\documentclass[11pt]{article}
\usepackage{amsmath}
\usepackage{amsthm}
\usepackage{mathrsfs}
\usepackage{amsfonts}
\usepackage{amssymb}
\usepackage{stmaryrd}
\usepackage{graphicx}

\usepackage{amssymb}
\usepackage{graphicx}
\usepackage{dcolumn}
\usepackage{bm}
\usepackage{subfigure}
\usepackage{rotating}
\usepackage{epsfig}
\usepackage{hyperref}
\newtheorem{theorem}{Theorem}[section]

\textwidth 16cm \textheight 23cm \graphicspath{{../images/}}

\date{}
\begin{document}
 \title{Nodal domain partition and the number of communities in networks
 \thanks{
This work is supported by the National Natural Science Foundation of
China (No.10971137), the National Basic Research Program(973) of
China (No.2006CB805900), and a grant of Science and Technology
Commission of Shanghai Municipality (STCSM, No.09XD1402500). }}
\author{ Bian He$^{1}$, Lei Gu$^1$,  Xiao-Dong Zhang$^{1}$ \\
\thanks{Corresponding author:\    Xiao-Dong Zhang, E-mail:
xiaodong@sjtu.edu.cn} $^1$
 Department of Mathematics, Shanghai Jiaotong  University,\\
           800 Dongchuan Road, Shanghai, 200240, P.R. China.\\
}
 \maketitle \thispagestyle{empty}

\begin{center}
\begin{minipage}{100mm}
Abstract: It is difficult to detect and evaluate the number of
communities in complex networks, especially when the situation
involves with an ambiguous boundary between the inner- and
inter-community densities. In this paper, Discrete Nodal Domain
Theory could be used to provide a criterion to determine how many
communities a network would have and how to partition these
communities by means of the topological structure and geometric
characterization. By capturing the signs of certain Laplacian
eigenvectors we can separate the network into several reasonable
clusters. The method leads to a fast and effective algorithm with
application to a variety of real networks data sets.


{\bf PACS classification codes:} 89.75.Hc, 05.10.-a, 02.70.Hm,
02.10.Ud.
\end{minipage}
\end{center}

\maketitle


\section{Introduction}
With the surveys of many real-world networks, including the
World-Wide Web \cite{AJB, Faloutsos}, metabolic networks
\cite{HBRZA,Fell and Wagner}, epidemiology \cite{Moore and
Newman,Pastor-Satorras and Vespignani}, scientific collaborations
and citation networks \cite{MEJN,Redner}, plenty of models were
proposed to study their topological features and dynamic behaviors,
such as the WS model \cite{Watts and Strogatz}, BA model \cite{Bara
and Albert} and Random Configuration model \cite{Chung2}. Recently,
a particular and useful network structure, which is called
``communities or clusters", has been appealed to considerable
attention. For its having no precise definition yet, common
description about communities is division of nodes into groups with
dense connection \emph{inside} and sparse connection \emph{outside},
by which mean communities could play an important role as the basic
functional components in the foundation of complex networks.

Practically, part of scientists concern with nodes' or edges'
individual behaviors affecting the surroundings \cite{Amaral and
Guimera, Freeman1977}, while some others focus on the dynamics of
ensembles of all the nodes in networks \cite{Menezes and Bara,
DALRSB}. These two special emphasis clarifies two possible
directions in community detection: (1)\emph{detail exploitation}:
identifying nodes or edges whose absence influences networks'
dynamic most, which mostly are the boundary parts; (2)\emph{global
sights} testing for partitions structure either close enough to the
original network or distinct most from the corresponding random
pattern network.

It has been long accepted that \emph{centrality measures}
\cite{Sabidussi, Freeman1979, Fortunato, Stephenson and Zelen} works
well in characterizing relative importance of a node in a network
\cite{Bonacich, Sabidussi, Granovetter}, as well as similar scales
on edges. In 2002, a fast algorithm aiming at identifying each edge
of the network a betweenness measure gave rise to a explosive growth
of activities in this field. Girvan and Newman \cite{Girvan} used
the scale to quantify edges' roles in the information transmission
following paths of minimal length across the network. Removal of
edges with high \emph{betweenness} could lead to an exposure of the
community structure.

Besides these \emph{detail exploitation} methods, scientists also
take the network whole  into account,  at which point two extreme
situations might be queried: how much the original network shifts
from its corresponding un-clustered version and spontaneously the
well-clustered one.

Consideration on the first situation educed one of the most popular
quality functions, so called \emph{modularity}
\cite{Newman2004-3,Newman2006, Newman2006-2}. The basic thought was
to compare the difference between total actual fractions of edges
inside groups and the expected fractions when edges were placed at
random. An improved version \cite{Newman2006} was also developed to
measure the difference between actual network and \emph{null model}
which yielded networks that were not supposed to have natural
community structures \cite{Molloy and Reed,Newman2006}.

Furthermore, a probabilistic framework embedded with `stochastic
matrices' provides another important quality function
\cite{eweinan-1, eweinan-2}. By introducing a metric on the space of
Markov chains $K$ which represent random walks on the network
\cite{eweinan-2}, simple stochastic structure could be finally
detected as the best approximation to the dynamics behaviors of $K$.
This simple structure may contain the community information of the
original network.

Despite the algorithm complexity of NP-hard \cite{Luxburg}, improved
approximation techniques in partition problems calculate `optimized'
partition under certain constraints \cite{Newman2004-2, Newman2006,
Jiang, eweinan-1}, however, the question about the number of
communities is still not easy to be answered \cite{Newman2006,
eweinan-2}.

In this paper we introduce a method applying weak-nodal-domain
partition(WNDP) which can suggest the number of clusters by
exploring the information in the  Laplacian eigenvectors. In Sec.II,
we begin with a brief review of the spectral partition method and
classical Nodal domain Theorem. In Sec.III, the main method and
algorithm will be presented respectively. In Sec.IV, the algorithm
is applied to three real network cases with fast and efficient
results. In Sec.V, exceptional case is demonstrated for the further
understanding of the method.

\section{Spectral Partition and Nodal Domains Theory}
The exactly mathematical definition of \emph{cluster} has not been
explicit so far, as though common agreement is focused on the
minimization of edges whose disappearance will separate the network
into groups with no inter-connection. A reasonable mathematical
framework should be required to grasp the essential properties of
the network not only precisely but also effectively. Particularly,
quite a number of the clustering methods are involved with special
matrices, for example, adjacency matrix, Graph Laplacian \emph{et
al}. All these matrices share the topological information of the
networks ostensibly or inconspicuously.

So far, surprising results have already been made clear that
eigenvectors of special matrices do work well in clustering
\cite{DMPRS,Merris}. Scientists are interested in the projection
from properties of these matrices to the corresponding networks'
outperforming structure. In the following sections, the eigenvectors
of the Graph Laplacian $\mathcal{L}$ will be served as a tool to
detect and analyze the community structure of the networks.


\subsection{Spectral Partition}

Let $G=(V,E)$ be an undirected graph with labeled vertex set
$V=\{1,2,\cdots,n\}$. As an unweighted graph, the \emph{adjacency
matrix} $A=(A_{ij})$ is defined to be
\begin{equation*}
A_{ij}= \left\{\begin{aligned}
        1, & & & \text{ if  $i \sim j$,} \\
        0, & & & \text{ otherwise.}
        \end{aligned} \right.
\end{equation*}
Meanwhile, the unnormalized graph Laplacian matrix is defined to be
$\mathcal{L}=D-A$. Here, $D=Diag(d_1,\cdots,d_n)$ is the
\emph{diagonal degree matrix} where $d_i=\Sigma_{j=1}^n A_{ij}$.
Another important concept is the \emph{cut size}:
\begin{equation}\label{equ:1}
Cut= \frac{1}{2}\sum_{\begin{subarray}{c} \text{$i,j$
in}\\\text{different}\\\text{group}
\end{subarray}} A_{ij},
 \end{equation}
Noting that the factor $\frac{1}{2}$ was a compensation for the
double count as $A_{ij}=A_{ji}$. It is the number of edges
connecting different communities. Traditional way is to minimize the
\emph{cut size} under all the possible partition choices.

Given a partition of $V$ into k sets $A_1,A_2,\cdots,A_k$, we
rewrite Eq.(\ref{equ:1}) to a universal form:
\begin{equation}\label{equ:2}
Cut(A_1,\cdots,A_k)= \frac{1}{2}\sum_{l=1}^k{Cut}_l,
\end{equation}
where ${Cut}_l=\Sigma_{i\in  A_l,j\in \overline{A}_l} A_{ij}$ and
$\overline{A}_l$ is the complement of $A_l$ in $V$.

We then set a $n\times k$ matrix $S=(S_{ij})$ to indicate the
positions of vertices in communities:
\begin{equation}\label{equ:3}
S_{ij}= \left\{\begin{aligned}
        1, & & & \text{ if  $i \in  A_j$,} \\
        0, & & & \text{ otherwise.}
        \end{aligned} \right.
\end{equation}
Note that the columns $S_j=(S_{1j},\cdots,S_{nj})^T$ of $S$ are
mutually orthogonal, and the matrix satisfies normalization
$\text{Tr}(S^TS)=n$. Thus,
\begin{eqnarray}\label{equ:4}
{Cut}_l = \sum_{i\in  A_l,j\in \overline{A}_l} A_{ij} = \sum_{i\in
A_l}\sum_{j} A_{ij}-\sum_{i\in  A_l}\sum_{j\in  A_l} A_{ij},
\end{eqnarray}
now put Eq.(\ref{equ:4}) into a matrix form
\begin{eqnarray*}
{Cut}_l &=& \sum_{i} D_iS_{il}S_{il}-\sum_{ij}A_{ij}S_{il}S_{jl}\nonumber\\
&=& S_l^T\mathcal{L}S_l,
\end{eqnarray*}
Hence,
\begin{eqnarray}\label{equ:5}
Cut(A_1,\cdots,A_k)&=&\frac{1}{2}\sum_{l}S_l^T\mathcal{L}S_l\nonumber\\
 &=&\frac{1}{2}\text{Tr}(S^T\mathcal{L}S).
\end{eqnarray}
So the problem of minimizing  $Cut(A_1,\cdots,A_k)$ can be rewritten
as
\begin{equation}\label{equ:6}
\frac{1}{2}\min_{\begin{subarray}{c}
A_1,\cdots,A_k\\\text{Tr}(S^TS)=n
\end{subarray}}\text{Tr}(S^T\mathcal{L}S) \text{  for all $S$ in Eq.(\ref{equ:3})}.
\end{equation}

Note that the optimization problem is based on matrix $S$ whose
entries' elements are only allowed to take special values 0 or 1,
which indicates complicated calculation in solving the issue. In
order to find a possible solution, we make a relaxation on the
discreteness condition and allow $S_{ij}$ to be arbitrary values in
$\mathbb{R}$. Hence, a general form is
\begin{equation}\label{equ:7}
\min_{\begin{subarray}{c} A_1,\cdots,A_k\\S^TS=I_{k\times k}
\end{subarray}}\text{Tr}(S^T\mathcal{L}S) .
\end{equation}

This is the standard form of a trace minimization problem, and the
Rayleigh-Ritz theorem tells us the best minimization can be achieved
by $S$ containing the first $k$ eigenvectors of $\mathcal{L}$ as its
columns. Let us make use of the Laplacian eigenvectors to rewrite
Eq.(\ref{equ:7}).

Since $\mathcal{L}$ is a positive semi-defined symmetric matrix, its
eigenvalues are all real and non-negative. Thus, respectively, let
$\lambda_1 \leq \lambda_2 \leq \cdots\leq \lambda_k\cdots
\leq\lambda_{n}$ be defined as the eigenvalues of $\mathcal{L}$, as
well as theirs corespondent normalized eigenvectors $f_{1},\cdots,
f_{n}$. For each row of Laplacian matrix $\mathcal{L}$, we have
\begin{equation*}
\sum_{j=1}^n\mathcal{L}_{ij}=D_{ii}-\sum_{j=1}^nA_{ij}=0.
\end{equation*}
This implies that $(1,\cdots,1)^T$ is the eigenvector of
$\mathcal{L}$ with eigenvalue $0$, so $0=\lambda_1 \leq \lambda_2
\leq \cdots\leq \lambda_k\cdots \leq\lambda_{n}$.

Thus, $\mathcal{L}=F\mathcal{D}F^T$, where the eigenvector matrix
$F=(f_{1}|\cdots|f_{n})$ and $\mathcal{D}$ is a diagonal matrix with
$\mathcal{D}_{ii}=\lambda_i$.
\begin{eqnarray}
\text{Tr}(S^T\mathcal{L}S)
&=&\sum_{l=1}^kS_l^T\mathcal{L}S_l\nonumber\\
&=&\sum_{j=2}^n\lambda_j\sum_{l=1}^k(f_j^TS_l)^2\label{equ:8}\\
&=&\sum_{j=2}^n\lambda_jw_j\label{equ:9}.
\end{eqnarray}
where $w_j=\sum_{l=1}^k(f_j^TS_l)^2$, by which means the
optimization has been split into pieces according to the eigenvalues
$\{\lambda_j\}$ as shown in Eq.(\ref{equ:8}) with the column-based
operation on the eigenvector matrix $F$. Our course would be clear:
to find a separation placing as much as possible of the weight $w_j$
on the side of small eigenvalues while as little as possible on the
large ones. In terms of Eq.(\ref{equ:9}), the properties of
eigenvectors are of great influence in the processing, which leads
to a further exploration into the eigenvectors space to help
approximating the optimization.

\subsection{Nodal Domains by Eigenvectors}
Recall the basic characterization of the eigenvalues and
eigenvectors
\begin{equation*}
\mathcal{L}f_j=\lambda_j f_j.
\end{equation*}
By multiplying both sides with vector $f_j^T$, we have
\begin{equation*}
f_j^T\mathcal{L}f_j=\lambda_j f_j^Tf_j=\lambda_j,
\end{equation*}
where $f_j$ is normalized and $\lambda_j$ is the eigenvalue in terms
of the Rayleigh quotient of $\mathcal{L}$. Actually, for $j\in
\{2,\cdots,n\}$, eigenvalues can be characterized as
\begin{equation}\label{equ:10}
\lambda_j=\inf_{f\perp W_{j-1}}f^T\mathcal{L}f,
\end{equation}
where $W_{j-1}$ is the subspace spanned by the eigenvectors of the
smallest $j-1$ eigenvalue \cite{Chung}. Simple calculation shows
that for arbitrary $f\in R^n$
\begin{equation*}
f^T\mathcal{L}f= \sum_{u\thicksim v}(f(u)-f(v))^2,
\end{equation*}
which can be substituted into Eq.(\ref{equ:10}) to form
\begin{equation}\label{equ:11}
\lambda_j=\inf_{f\perp W_{j-1}}\sum_{u\thicksim v}(f(u)-f(v))^2,
\end{equation}
where $\lambda_j$ is achieved with $f$ being the exact $j$-th
corresponding eigenvector.

It is easy to see that under condition $f\perp W_{j-1}$, the
eigenvector $f_j$ is the weight function that minimizes the total
weight difference between pairs of adjacent nodes through the whole
network. Such a property indicates a structure in which nodes are
more likely to appear in the same group if their corresponding
elements in $f_j$ are close \cite{Luxburg}. However, these values
are widely distributed that there is no such a criteria to depict
the boundary between groups.

\begin{figure}[htbp]
\begin{center}
\subfigure[]{
\includegraphics[width=3in]{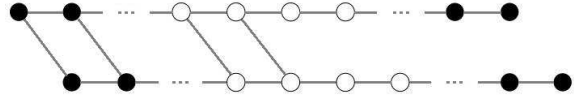}}
\caption{\label{fig:sample} The cockroach graph from Guattery and
Miller \cite{Guattery and Miller} has 80 nodes as illustrated above
with each suspension points representing a line of 16 nodes. Here,
we apply the fourth eigenvector to get this special partition where
black for nodes of negative eigenvector values, white for positive
ones. It's easy to see that $w_4$ reaches its maximum when nodes
with the same signs are in the same groups, while the graph
structure tells us negative nodes are divided into three unconnected
parts.}
\end{center}
\end{figure}

Let's look back upon Eq.(\ref{equ:9}). Matrix $S$, which represents
the partition, should be chosen to make $w_j$ relatively large while
$j$ is small. This idea could offer a possible way to describe the
criteria. For a single $f_j$, the corresponding $w_j$ reaches the
maximization when nodes are grouped as follow:
\begin{equation}\label{equ:12}
 \left\{\begin{aligned}
        x\in S_1, & & & \text{ if  $f_j(x)>0$,} \\
        x\in S_2, & & & \text{ if  $f_j(x)<0$.}
        \end{aligned} \right.
\end{equation}
Noting that set $\{x~|f_j(x)=0\}$ is not mentioned for its making no
contribution to the maximization of $w_j$. However, the network
structure is complicated that nodes in the same group $S_1$ or $S_2$
might not be connected at all. See Fig.\ref{fig:sample} for
instance.

Hence, instead of clustering nodes as in Eq.(\ref{equ:12}), we add a
natural constraint that each group should be a connected subgraph
referring to the graph geometry structure. The square operation in
Eq.(\ref{equ:8}) suggests that two operations on subgraphs are
preferred: (a) nodes within the same subgroup share the same sign in
$f_j$; (b) each subgroup contains as many nodes as possible. This
consideration leads us to focus on partitions deduced according to
the signs of each eigenvectors' elements.

An interesting result named the Courant-Hilbert nodal theorem
\cite{Courant and Hilbert} gives a detail description about the
domains cut by zeros of each eigenfunctions.

{\small{\it Given the self-adjoint second order differential
equation $L[u]+\lambda\rho u = 0$ ($\rho > 0$) for a domain G with
arbitrary homogeneous boundary conditions, if its eigenfunctions are
ordered according to increasing eigenvalues, then the zeros of the
$n$-th eigenfunction $u_n$ divide the domain into no more than $n$
subdomains. No assumptions are made about the number of independent
variables.} [From Courant-Hilbert: Methods in Mathematical
Physics.]}

It inspires us to take into account the corresponding discrete
situation, which is the discrete nodal domain theorem on graph
$G=(V,E)$. We define a positive (negative) strong nodal domain of a
function $f$ on $V (G)$ to be a maximal connected induced subgraph
of $G$ on vertices $v \in V$ with $f(v) > 0$ $(f(v) < 0)$.
Meanwhile, a positive (negative) weak nodal domain of a function $f$
on $V (G)$ is a maximal connected induced subgraph with $f(v) \ge 0$
($f(v) \le 0$) that contains at least one nonzero valued node. The
relation between eigenvectors and nodal domains was introduced and
studied by Davies, Gladwell, Leydold, Stadler, \emph{et al.}
\cite{Biyikoglu, Dekel Lee Linial}. One of those important results
is

\begin{theorem}
\emph{\cite{Biyikoglu}} Let $M$ is a symmetric matrix with
nonnegative diagonal entries and $M_{uv}<0$ as $u\sim v$. Let
$\lambda_1 \leq \cdots <\lambda_k=\lambda_{k+1}
\cdots=\lambda_{k+r-1}<\lambda_{k+r}\leq\cdots\leq\lambda_n$ be
eigenvalues of $M$ and $f_k$ be the corresponding eigenvector of
$\lambda_k$. Then, the number of strong nodal domains of $f_k$ is no
more than $k+r-1$, and the number of weak nodal domains of $f_k$ is
no more than $k$.
\end{theorem}

This theorem defines the matrix $M$ in a general expression, in
which the Laplacian of graph is included. It shows a possible
natural structure framework in which we achieve a maximization of
$w_j$ satisfying the connectivity information. An interesting
viewpoint occurs when there are zero elements in the eigenvectors.
Weak nodal domains request a sharing of these nodes, which means an
overlapping structure might be naturally defined, or certain basic
functional parts in the dynamic system of the network are
discovered.

\section{Partition by weak nodal domain}
\begin{figure}[tbp]
\begin{center}
\subfigure[]{
\includegraphics[width=3in]{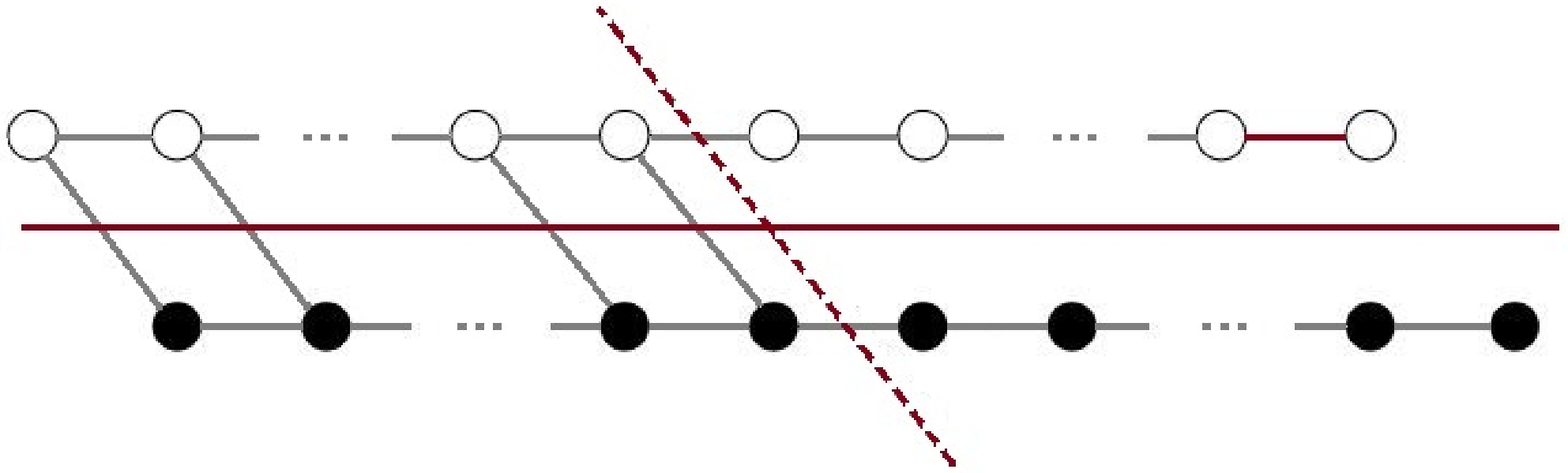}}
\subfigure[]{
\includegraphics[width=3in]{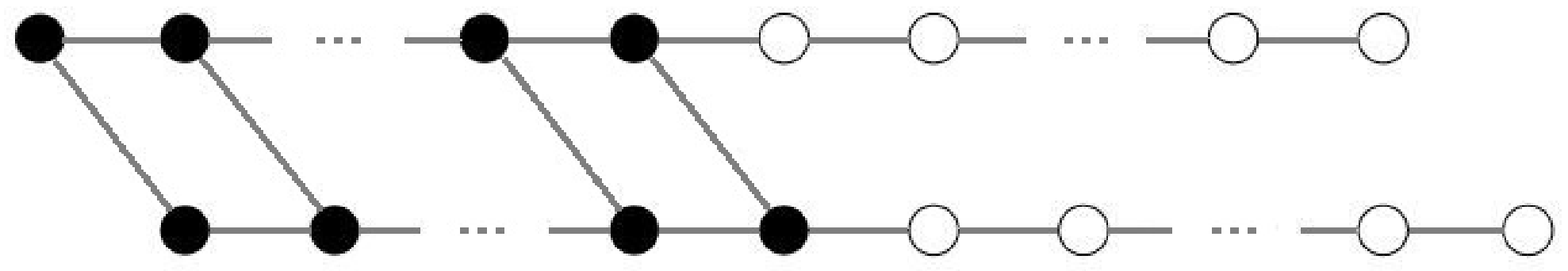}}
\caption{\label{fig:cock}The cockroach graph from Guattery and
Miller\cite{Guattery and Miller}, where black for negative
eigenvector nodes, white for positive eigenvector nodes. (a) the
horizontal cutting is deduced by the second eigenvector, the dashes
line implies the ideal separation; (b) partition deduced by the
third eigenvector.}
\end{center}
\end{figure}

Specially, eigenvector affording the second smallest eigenvalue of
Laplacian of the network is called the \emph{Fiedler vector}
\cite{Fiedler}. It has been applied in graph bi-partitioning
\cite{Pothen} and spectral clustering \cite{Belkin and Niyogi}.
Generally speaking, these methods provide partitions that attract
rather large weight on the smallest nonzero eigenvalue in
Eq.(\ref{equ:9}) to make $Cut(A_1,\cdots,A_k)$ small. However, bad
results can be found upon many graphs in \cite{Guattery and Miller}.
Take ``cockroach graph'' \cite{Guattery and Miller} for example, the
second eigenvector offers a division horizontally cutting through
the ladder while, obviously, the ideal cut is the long dots line
(see Fig.\ref{fig:cock}.(a)).

The discrete nodal domain theorem indicates that the \emph{Fiedler
vector} decides weak nodal domains no more than two, but for the
cockroach graph, the ideal cut separates the graph into three parts.
Spontaneously, we are interested in the behavior of the third
eigenvector. Calculation shows that it provides the exact separation
just as the ideal cut does (see Fig.\ref{fig:cock}.(b)).

This result leads us back upon Eq.(\ref{equ:9}). Despite the weight
expression $w_j=\sum_{l=1}^k(f_j^TS_l)^2$, eigenvalues also play an
important role in the minimization of the formula. Actually
cockroach graph possesses two eigenvalues $\lambda_2=0.0057$ and
$\lambda_3=0.0062$ that are relatively close, compared with
$\lambda_4=0.0246$ and other eigenvalues, which implies that heavy
weight on $\lambda_3$ is also a possible optimized choice. Hence, we
can apply the eigenvectors other than the \emph{Fiedler vector} to
partition and this is so called the weak-nodal-domain
partition(WNDP).

In most situations eigenvectors corresponding to smaller eigenvalues
provide WNDP with smaller cut size, which makes multi-communities
structure uneasy to be determined. To avoid this shortage, we take
the famous quality function modularity $Q$ \cite{Newman2006} as a
criterion.
\begin{equation*}
Q=\text{Tr}(S^T(A-P)S) .
\end{equation*}
where $P$=($P_{ij}$) is the random configuration correspondence of
$A$ satisfying $P_{ij}=\frac{d_{i}d_{j}}{2|E|}$ \cite{Newman2006}.

This quality function depicts the difference between how many edges
within communities and how many edges to be expected within
communities. Interestingly, we have
\begin{eqnarray*}
Q &=& \text{Tr}(S^T((D-P)-(D-A))S) \\
  &=& \text{Tr}(S^T(D-P)S)-\text{Tr}(S^T(D-A)S),
\end{eqnarray*}
where the first half represents expected edge number outside
communities and the other half dedicates actual edge number of such
set. Note that the second half is exact the \emph{cut size} of the
network mentioned before. We are interested in the difference
between WNDP and the exact optimization result of modularity $Q$.

Consider a WNDP, $Q$ will be improved in three ways:

(i)two communities merge together: volume to be $V_i=\sum_{h\in
A_i}d_h$ and $V_j=\sum_{h\in A_j}d_h$, number of connecting edges to
be $c_{ij}$, this alteration only happens when
\begin{equation*}
\frac{V_i*V_j}{2|E|}<c_{ij};
\end{equation*}

(ii)one community split into two: with the same definition as above,
this alteration only happens when
\begin{equation*}
\frac{V_i*V_j}{2|E|}>c_{ij};
\end{equation*}

(iii)move single vertex from one community to another: as vertex v
with $n_i$ connections to $A_i$ and $n_j$ connections to $A_j$, it
belongs to $A_i$ to make better benefit on $Q$ when
\begin{equation*}
\frac{V_i*n_i}{2|E|}-\frac{V_j*n_j}{2|E|}<n_i-n_j.
\end{equation*}

The first two operations usually will not be needed if the community
structure are relatively clear (see samples below), thus slight
alterations on vertices may lead to a good approximation.

%
%
%

This algorithm is practically under processing in the form of matrix
calculation. Specially, given the function $f$, the complexity of
finding the weak nodal domains is $O(|V|)$.

Remember that there could be zero elements in the eigenvectors.
Define a vertex $v$ to be a zero vertex if $f(v)=0$. Similarly a
zero component is a maximal connected subgraph of zero vertices. We
preprocess the graph in following steps:

(i)Contract all zero components into single vertices which inherit
the connection of the components. Here, multiple edges are allowed.
This leads to graph $G_1$.

(ii)Split each zero vertex in $G_1$, say $v$, into two connected
individual vertices $v^{+}$ and $v^{-}$, in which $v^{+}$ inherits
all neighbors of v with positive values in $f$ and $v^{-}$ inherits
corresponding negative parts. (note that either all neighbors of v
are zero vertices or all v has neighbors of different signs). This
leads to graph $G_2$.

We call the new graph $G_2$ after the two steps a weak domain graph.
It is the graph our algorithm be applied on. Note that the most
complex part of the method is to calculate the eigenvectors of the
sparse Laplacian matrix, which is known to be polynomial of order
$O(n^2)$, and one can apply shifted power method to get the the k-th
eigenvector which is of complexity
$O(\frac{n}{\log(\lambda_k-\lambda_{k-1})})$.

Situation with zero components could be complicated, for that
modularity is defined on structure without overlapping. In this
paper, a limitation of the definition on zero value to be of order
$O(10^{-17})$ will avoid overlapping structure. The ambiguous
boundaries will be discussed in our further work.

\begin{figure}[b]
\begin{center}
\subfigure[Modularities of WNDPs by different eigenvectors of
Laplacian on Dolphin graph. WNDP by $f_2$ is the best choice.]{
\includegraphics[width=3in]{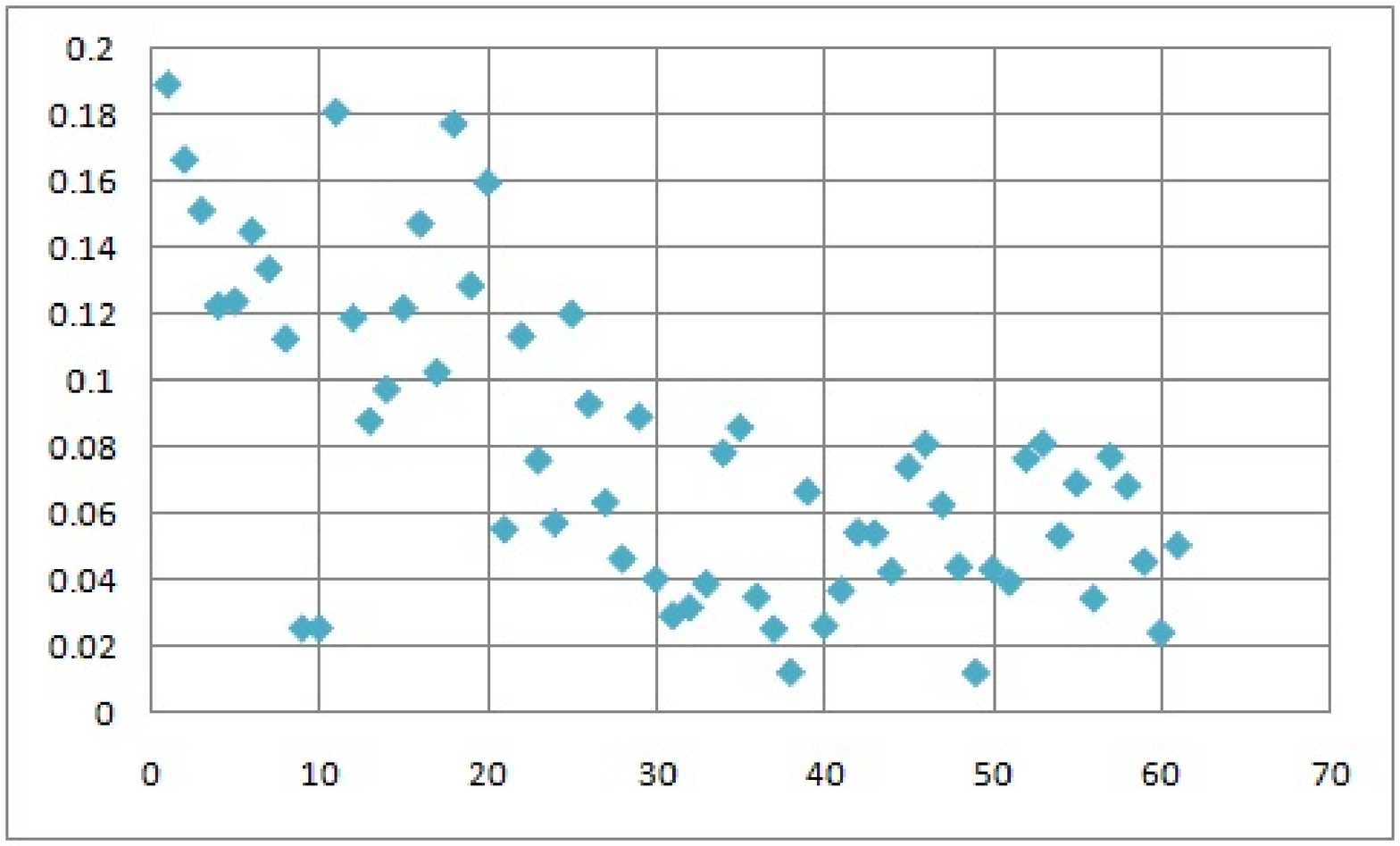}}
\subfigure[The colors represent real observed division of 62
bottlenose dolphins. The solid line represents the result of WNDP.
Follow our algorithm comes a partition with only one controversial
node.]{
\includegraphics[width=3in]{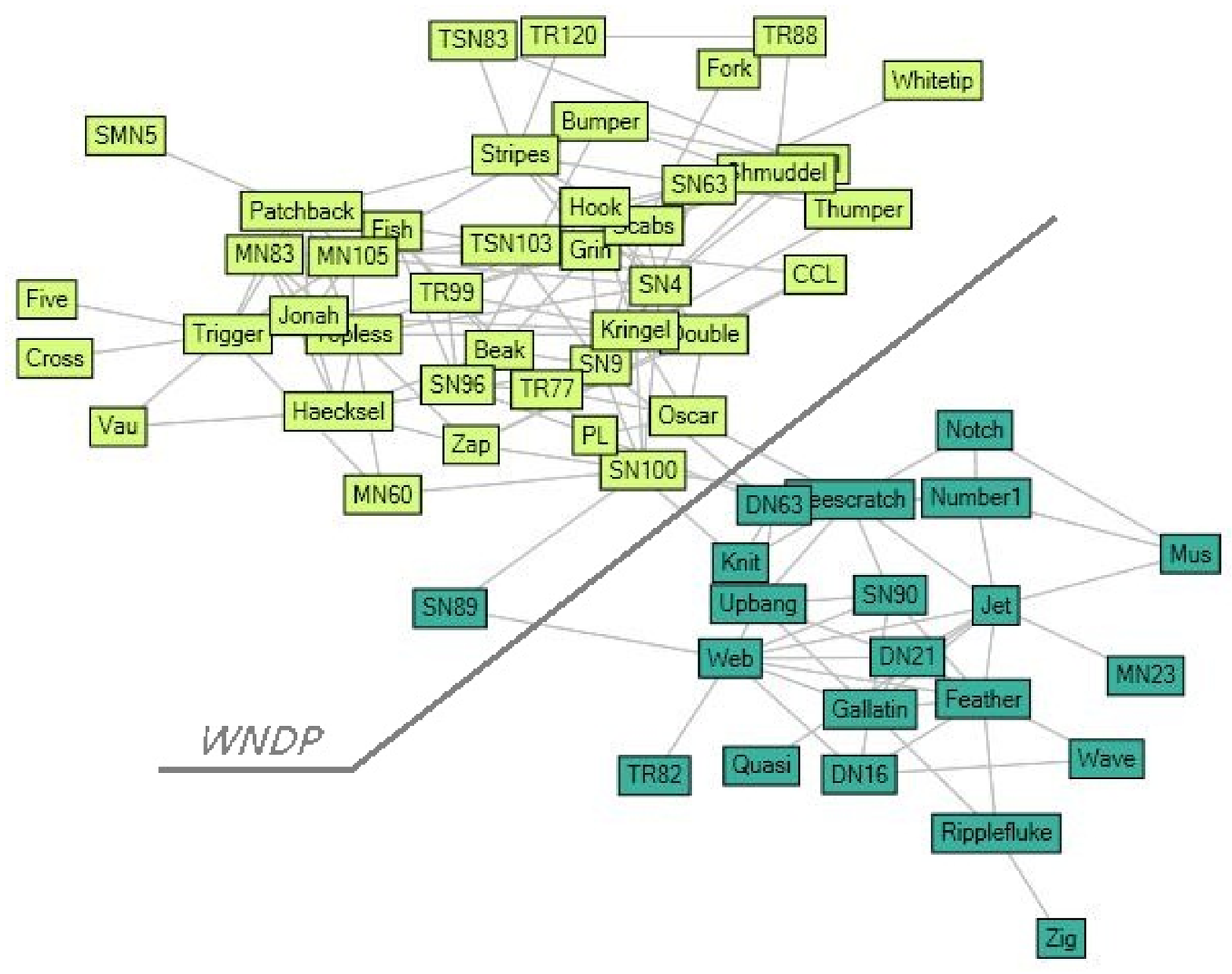}}
\caption{\label{fig:dol}The Dolphin Graph \cite{Lusseau}}
\end{center}
\end{figure}

\section{Experiments with WNDP}
The whole comparison framework define above allows us to test the
best partition as well as reasonable number of clusters. Or more
precisely, we are trying to find out what information the
eigenvectors might suggest for a possible partition. Here, three
different networks are conducted under our algorithm to test the
WNDP method.

\subsection{Dolphin Graph}
The establishment of Dolphin graph \cite{Lusseau} is based on
certain group of bottlenose dolphins living in Doubtful Sound, New
Zealand. The pattern that two dolphins getting alone together
suggests certain relation among members of this representative
animal social network. See Fig.\ref{fig:dol}(b) for the detail
connections. Scientists found an interesting phenomenon after years
of observation: the whole group of dolphins split into two small
subgroups following the departure of one key member named 'SN89'.
This observed division is represented with colors in the figure.

\begin{figure}[b]
\begin{center}
\subfigure[Modularities of WNDPs by different eigenvectors of
Laplacian on Political Books graph. WNDP by $f_2$ is selected by the
algorithm.]{
\includegraphics[width=3in]{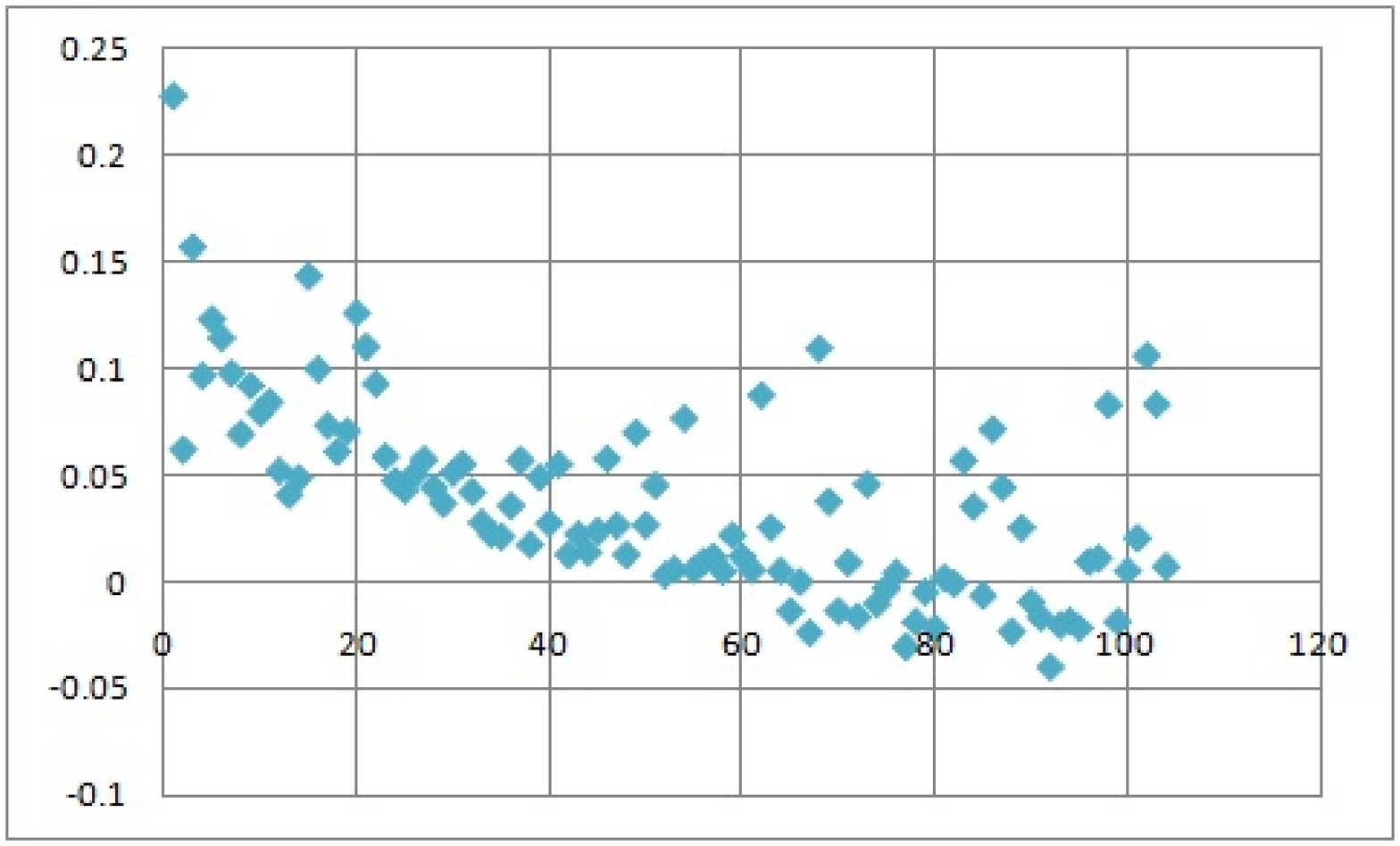}}
\subfigure[ Books with dark color are liberal, the grey ones are
centrist or unaligned, and the rests without colors are
conservative. The grey books act as buffers between the ones with
left-wing and right-wing points of view. Solid line divides the
graph following WNDP of $f_2$. Only two confused nodes representing
books purchased frequently by both sides are positioned
incorrectly.]{
\includegraphics[width=3.5in]{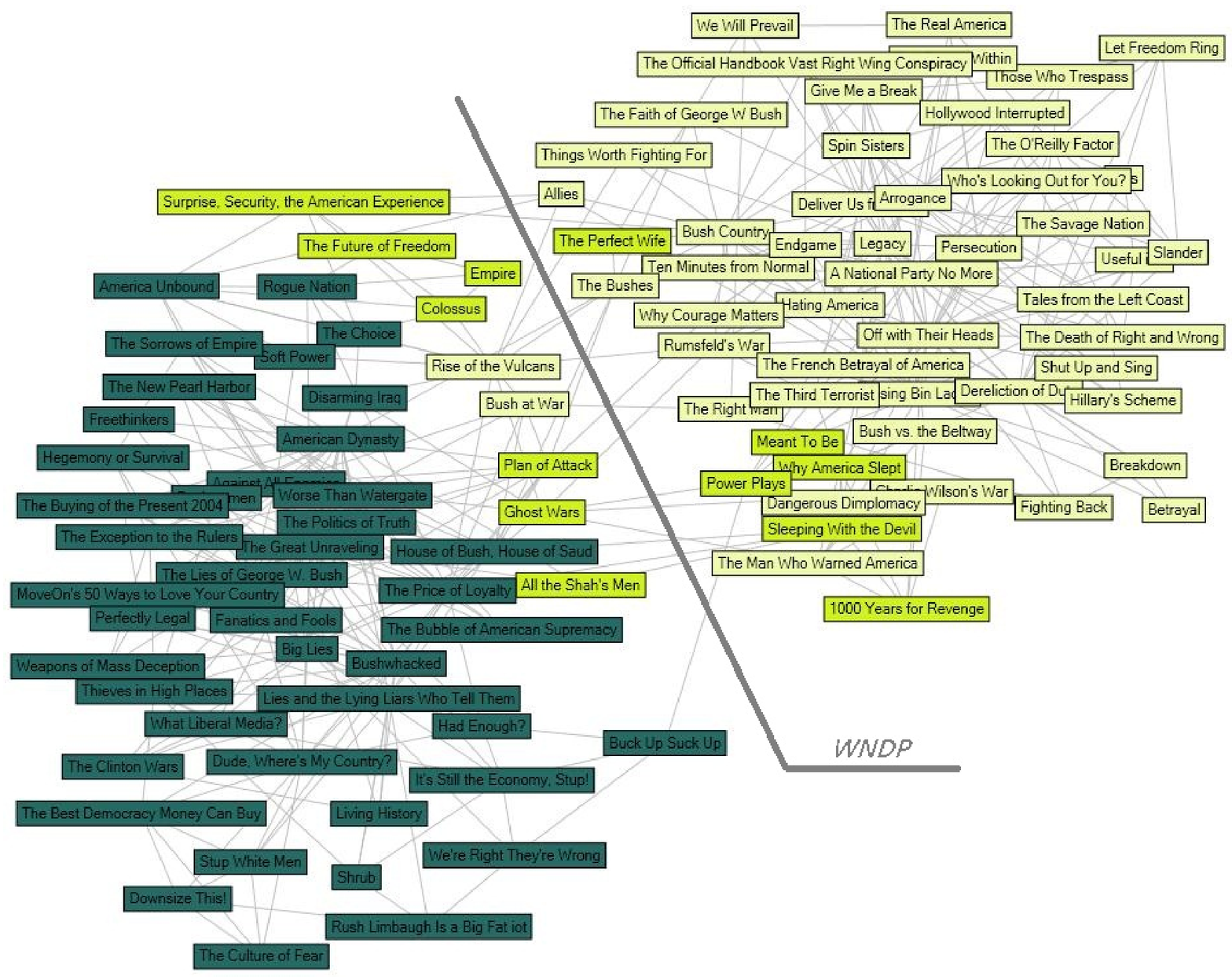}}
\caption{\label{polbook}The Political Books graph \cite{Krebs}}
\end{center}
\end{figure}

We process the graph with our algorithm, and the outcome indicates
$f_{2}$ possessing reasonable WNDP(see Fig.\ref{fig:dol}(a)). Solid
line in Fig.\ref{fig:dol} depicts the division of the calculation
result. Only 'SN89' does not fit the real statement. The background
of this division tells us 'SN89' is an ambiguous node for its role
as a association between two subgroups. Thus, the result of our
method is quite close to the reality.

\subsection{Political Books Graph}

Political Books graph is assembled and studied by Krebs
\cite{Krebs}. The database was taken from the online bookseller
www.amazon.com, in which 105 books have been considered. Edges
between books represent frequent co-purchasing by the same buyers.
Krebs collected these information in 2004 around the US Election. He
wanted to study the relation between books describing different
political views. Naturally, this special graph inherits bilateral
structure arising from the two different political tendency in the
United States, democratic and republic respectively.

It is not surprising that WNDP by $f_2$ of Political Books graph
reaches its optimum modularity. Fig.\ref{polbook}(b) is the exact
partition, in which each side has its particular groups of authors
and readers. An intuitive survey of the original graph shows our
prediction still has two books of liberal presented in the group of
conservative.

\subsection{Capocci Graph}
\begin{figure}
\subfigure[\emph{up-left}: the original graph; \emph{up-right}:
partition by $f_2$; \emph{bottom-left}: partition by $f_3$;
\emph{bottom-right}: partition by $f_4$. Generally speaking,
eigenvectors other than these four above would possess more
complicated nodal domain structures.]{
\includegraphics[width=3in]{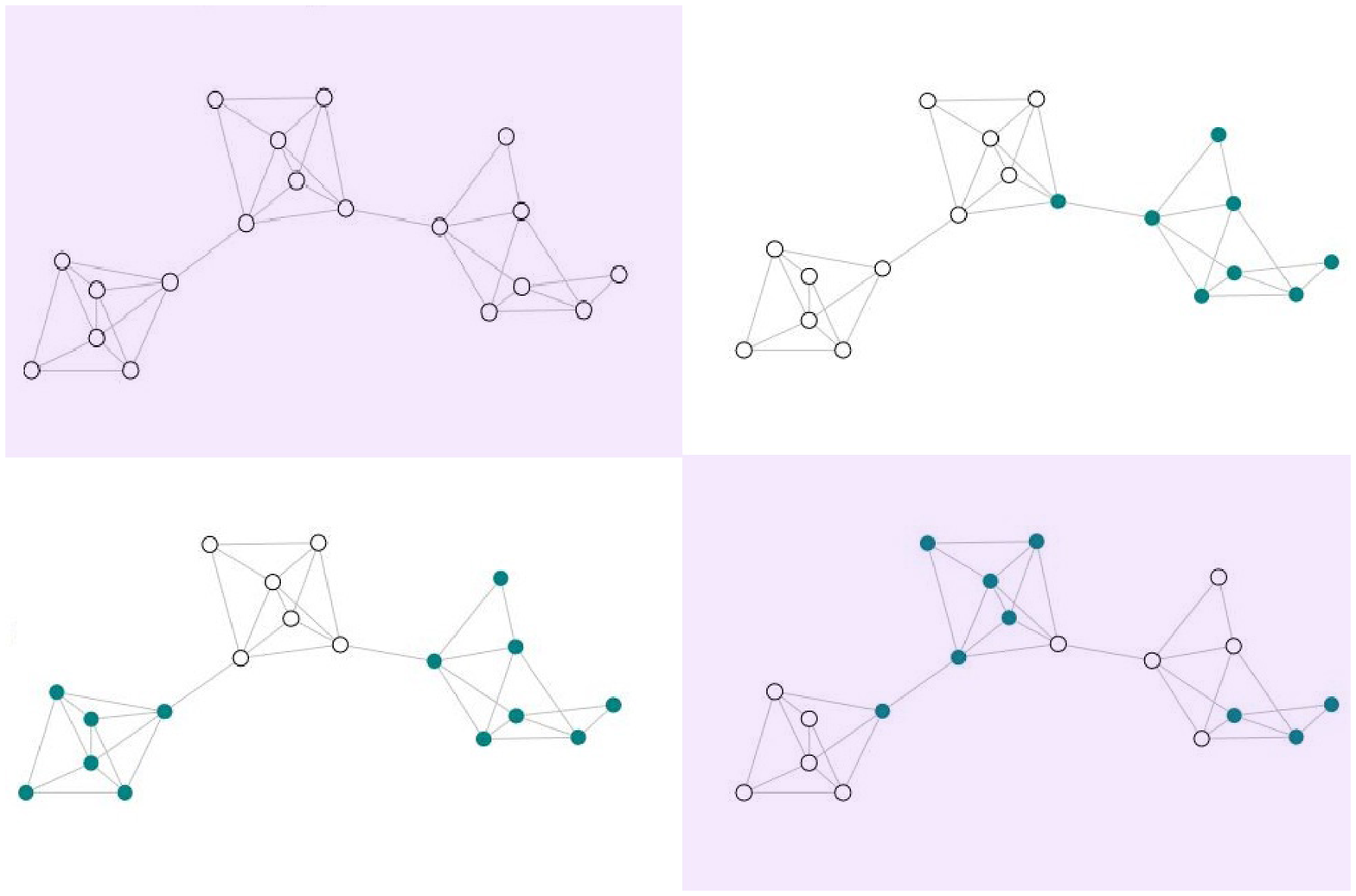}}
\subfigure[The eigenvalues of the Laplacian on Political Books
graph. Note that the first three eigenvalues are rather smaller than
the rests.]{
\includegraphics[width=3in]{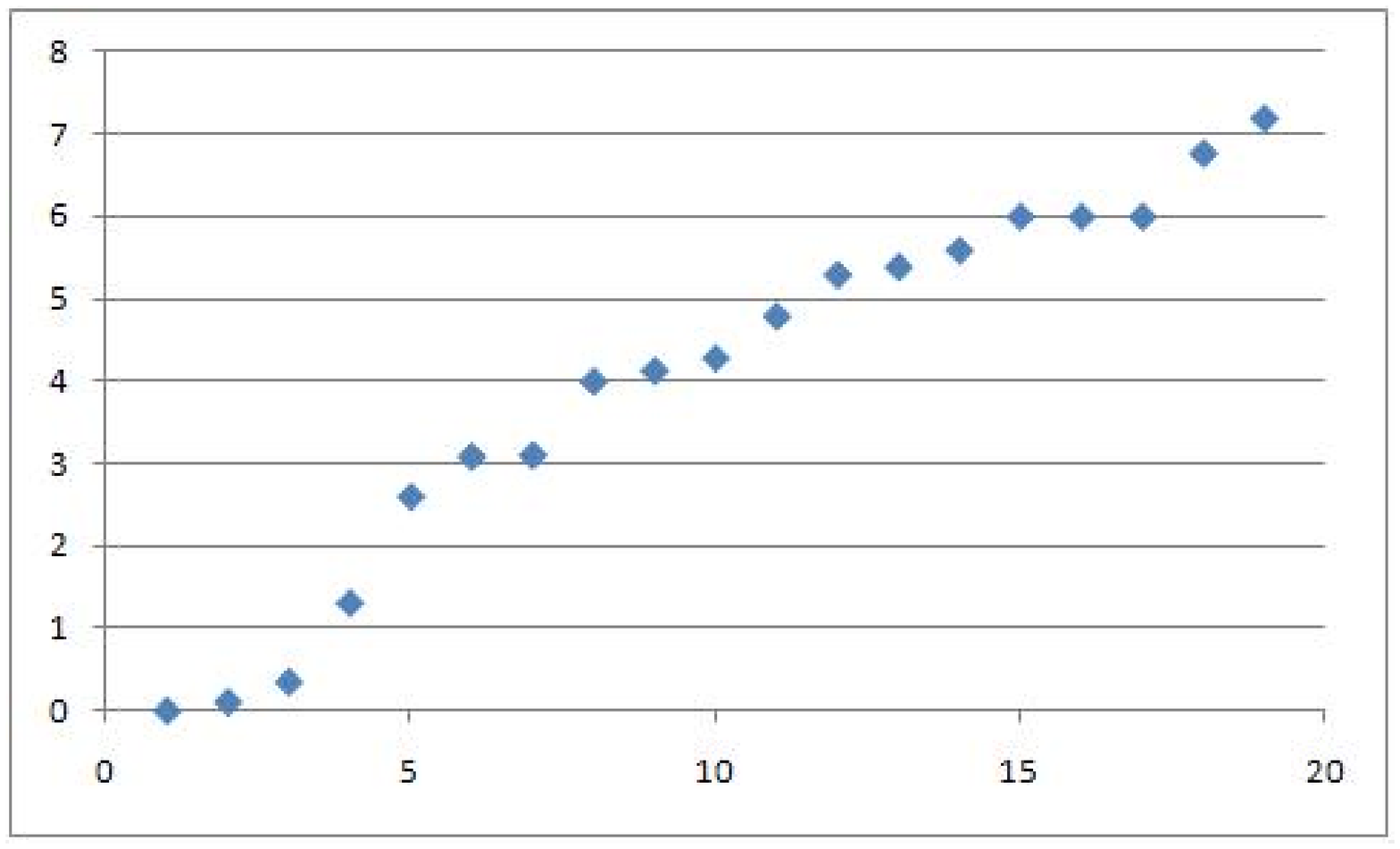}}
\caption{\label{fig:capocci}The Capocci Graph \cite{Capocci}}
\end{figure}

Capocci Graph is a simple graph (as shown in Fig.\ref{fig:capocci}
\emph{up-left}) generated by Capocci for the application of
eigenvector component to identify communities \cite{Capocci}. The
experiment shows that the second eigenvector of the right stochastic
matrix, which is $D^{-1}A$, indicates three plateaus corresponding
to the three evident component of the graph.

\begin{figure}[h]
\includegraphics[width=3in]{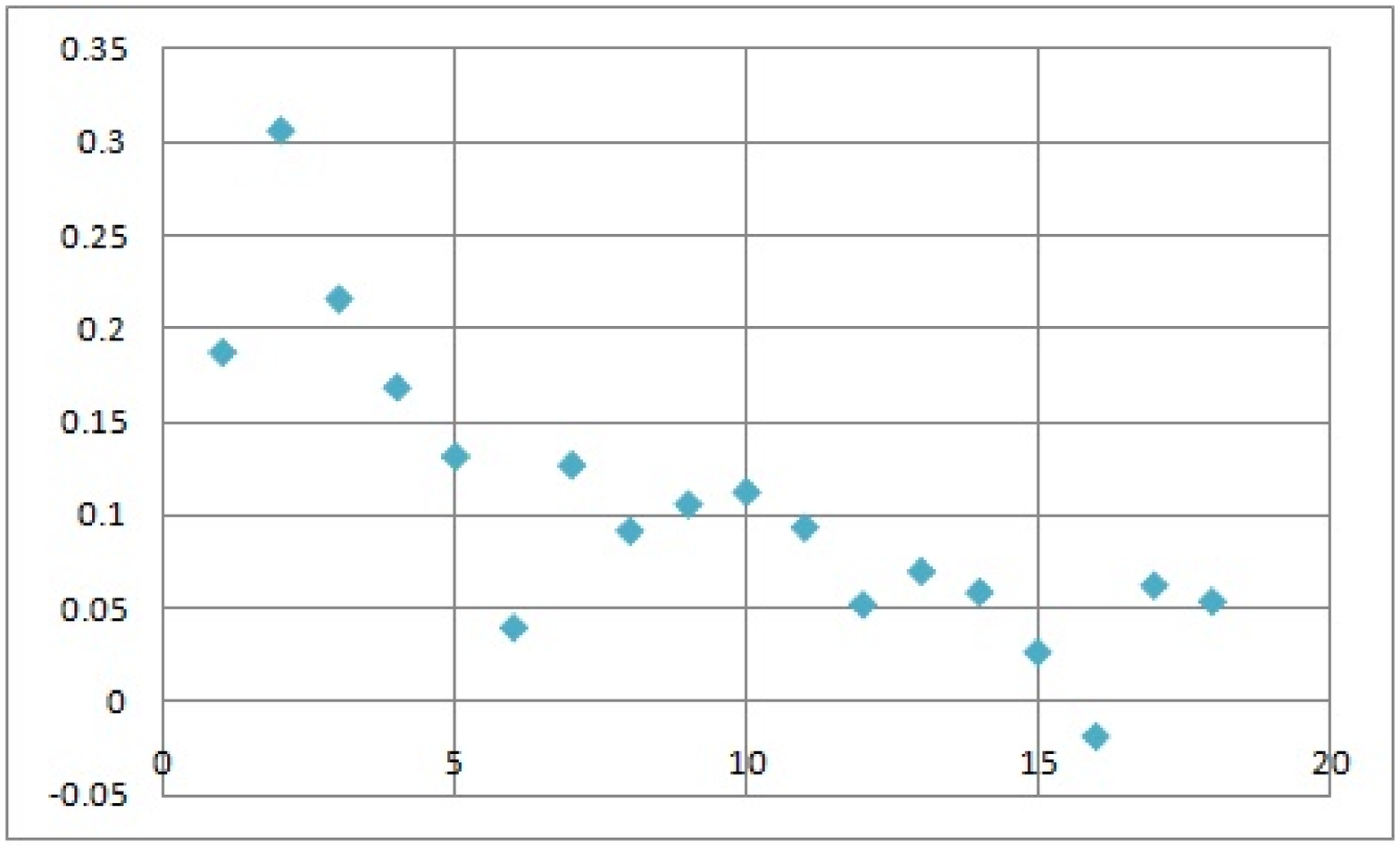}
\caption{\label{fig:capocci_m}Modularities of WNDPs by different
eigenvectors of Laplacian on Capocci graph, among which $f_3$
suggests the best choice.}
\end{figure}

Fig.\ref{fig:capocci_m} depicts the numerical result of our
algorithm that the largest modularity is offered by $f_3$. Detail
partition is represented in Fig.\ref{fig:capocci}(a), in which we
also demonstrate respectively the WNDPs of eigenvectors
corresponding to the first three nonzero eigenvalues. Obviously,
WNDP on the \emph{bottom-left} by the chosen $f_{3}$ gives a
partition matching our visual observation.

Compared with Dolphin Graph and Political Books Graph, the best WNDP
of Capocci Graph is the one according to $f_3$ other than $f_2$.
This small alternation reminds us of the important roles that
eigenvalues $\{\lambda_j\}$ have played in Eq.(\ref{equ:9}). To
illustrate this idea, we separate small eigenvalues with relatively
large ones
\begin{equation*}
\sum_{j=2}^n\lambda_jw_j=\sum_{j=2}^s\lambda_jw_j+\sum_{j=s+1}^n\lambda_jw_j.
\end{equation*}
where $\{\lambda_i\}_{i=1}^s$ represent relatively small eigenvalues
while $\{\lambda_i\}_{i=s+1}^n$ the large ones. Our experiments
suggest that eigenvectors corresponding to $\{\lambda_i\}_{i=1}^s$
may hold the information about the community structure of the graph.

\subsection{Computer-Generated Graphs}
\begin{figure}[b]

\includegraphics[width=3.5in]{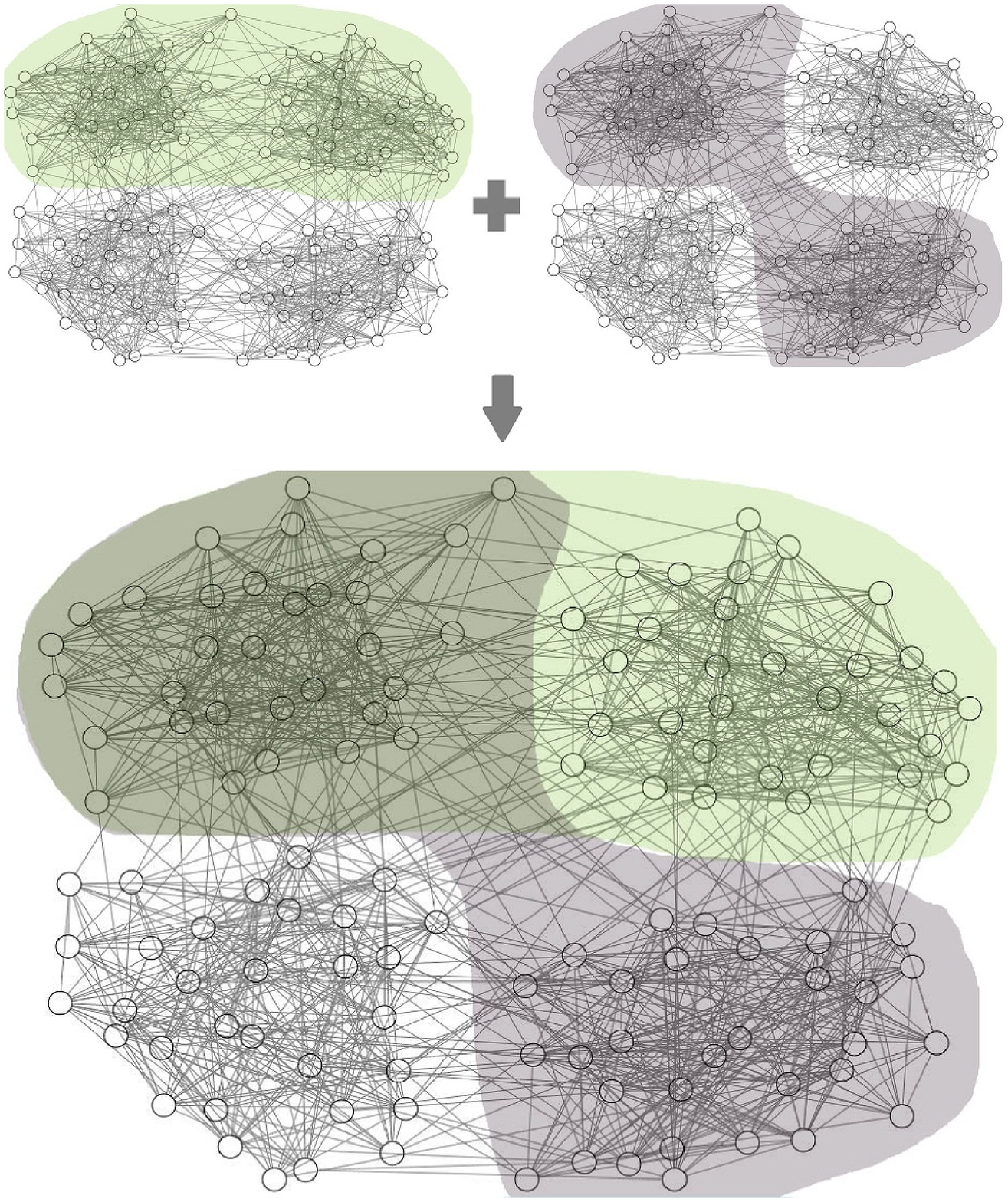}
\caption{\label{fig:artificial}The artificial sample assembled by
computer. The four communities are easily recognized through direct
observation. The color shade represents different clusters which are
suggested by the eigenvectors. Small graphs on the top are separated
by WNDPs corresponding to the second and fourth eigenvectors
respectively, in which only two clusters each are detected.
Combination of these two results reveals the exact community
structure, and they are exact the mathematical outcome of
generalized WNDP.}

\end{figure}

Take a rough look at the community structure, we contract each
cluster of the WNDP into one single node inheriting connections of
the cluster. Despite multiple edges, no circles exit in this simple
structure which, in brief, is a tree. This interesting phenomenon
comes from the fact that only two signs `+` and `-` are used to
identify different nodes. We generate a graph artificially following
the method used by Girvan and Newman \cite{Girvan}, also called the
ad hoc network, in which a whole graph of 128 vertices is divided
into four communities of 32 vertices each. More precisely, in our
special case, edges between pairs of nodes in the same community are
placed with possibility 0.4 while ones in different communities
share possibility 0.1. The randomness of the edges indicates
multiple choices, however, we are only interested in the known
community structure of these graphs. Graph in
Fig.\ref{fig:artificial} is one of these special samples following
the Girvan and Newman's method on which our algorithm is applied on.
It is easy to find out that at least four signs are required to
separate these four communities, which suggests our method being
incomplete.

Calculation shows that WNDPs by eigenvectors according to the first
three nonzero eigenvalues divide the graph into two parts
respectively (see Fig.\ref{fig:artificial}). In other words, each
eigenvector reveals part of the community structure. By combining
these partial information together, we may recover the acknowledge
of the whole structure. This leads us to a generalized definition of
the nodal domains: a strong nodal domain of functions
$\{f_1,\cdots,f_i\}$ on $V (G)$ is a maximal connected induced
subgraph of $G$ on vertices $v \in V$ which have corresponding
vectors $(f_1(v),\cdots,f_i(v))$ belonging to the same quadrant of
the i-dimension Euclid space. Note that 'i' is the exact number of
eigenvectors $\{f_1,\cdots,f_i\}$ on which we process for the
combined information. So is the alteration about the definition of
weak nodal domains.

Convictive result comes out by applying our algorithm on these
generalized WNDP (see Fig.\ref{fig:artificial}). Single eigenvector
does separate the whole network into reasonable parts, but not
subtly enough. Structure functioning with a relatively small scale
will be demonstrated by appropriate blend of certain eigenvectors.

\section{Conclusion}

The experiments above confirm that the nodal domains of the
Laplacian eigenvectors do hold certain information about the
community structure of the graph. Still, exception would arise when
the situation comes with relatively large difference between the
volume numbers of communities. Virtually, we set up a graph with
following three steps (see Fig.\ref{antisam}):
\begin{itemize}
\item
Use ER model \cite{ER} to build a graph, named $G_1$, of 200 nodes
whose average degree is 40 (nodes shaped in solid squares).

\item
Use ER model to build another graph, named $G_2$, of 20 nodes whose
average degree is 4 (nodes shaped in solid circles).

\begin{figure}[h]
\begin{center}
\includegraphics[width=3.5in]{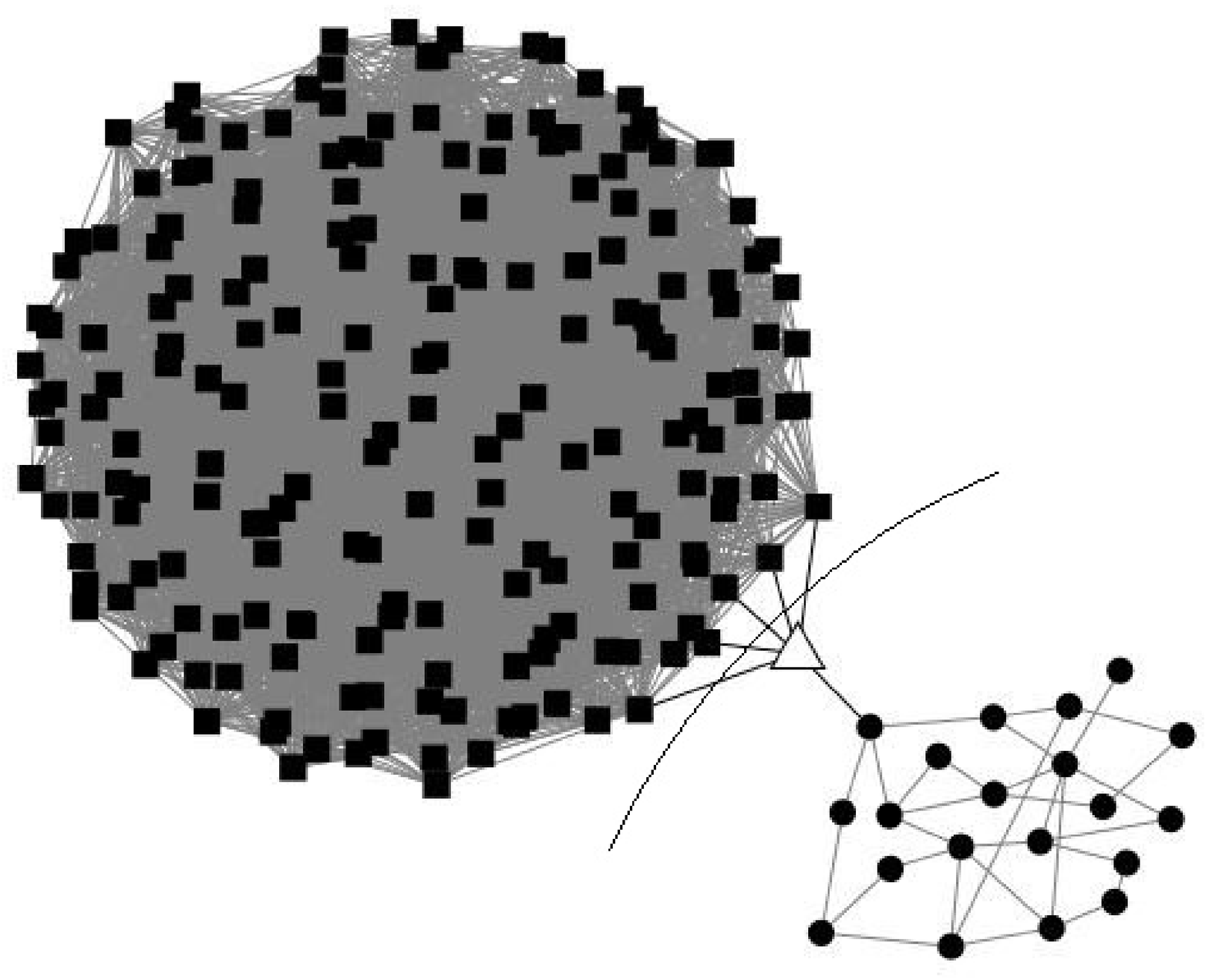}
\caption{\label{antisam} This special graph is established by
following the three steps above. The triangle in the figure is the
key contradiction. Solid line divides the graph by WNDP while direct
observation shows the triangle is placed incorrectly.}
\end{center}
\end{figure}
\item
Build an extra single node, first connect it with a random node in
$G_2$, then connect it with five random nodes in $G_1$ (node shaped
in triangle).
\end{itemize}

Calculation suggests that WNDP by the second eigenvector possesses
the largest modularity. Solid line in Fig.\ref{antisam} shows the
exact division. But direct observation tells us the triangle has
more connection with $G_1$ than that with $G_2$.

A survey into the algorithm knowledged us that the process of
calculating the eigenvectors is equal to evaluate each point with a
average of its neighbor's eigenvector value, but under a certain
rate which is the eigenvalue. Thus, the eigenvector $f_2$ of the
Laplacian on the graph would evaluate nodes in $G_1$ with positive
values while negative ones for $G_2$. Because $G_2$ has much less
nodes than $G_1$, $f_2$ on $G_2$ possess relatively small negative
values, which indicates that under the mean of average the triangle
could have a negative value as nodes in $G_2$ do. Further work
should be focused on the position amendment of the boundary nodes
like the triangle.

In section III, we already find this situation a solution by
checking the boundary vertices for improvement on modularity $Q$.
But recall the first two operations in modifying $Q$, we note that
if the volume of the two communities is small or unbalanced(with
wide gap), these two should not be separated for the inequality
being unsatisfied. That is why the modularity has a well-known
resolution limit, that makes clusters smaller than a given size
undetectable. A resolution coefficient $\lambda$ is welcome to make
up this limitation as the inequality goes as
\begin{equation*}
\frac{V_i*V_j}{2\lambda |E|}<c_{ij},  \lambda \in (0,1]
\end{equation*}
where the vertexes are localized that we only consider the influence
form their ranged neighbours.

 To conclude, in this paper we introduce the method that
applies weak nodal domains according to eigenvectors of the
Laplacian on the graph. We calculate modularity \cite{Newman2004-2,
Newman2006} to decide which WNDP behaves best that it may suggest
the optimal number of communities in networks. We also test our
algorithm in real-world models. As the examples show, for arbitrary
graphs, WNDP works quite well in deciding the number of clusters in
the graphs. For un-weighted graph the mechanism why the WNDP works
so good is still unknown, and we will work on this in our future
study as well as finding other good parameters to choose the best
eigenvectors.

\end{document}